\documentclass[aps,pra,twocolumn,superscriptaddress,showpacs,amsmath,amssymb,
floatfix]{revtex4}
\usepackage{graphicx}
\usepackage{hyperref}
\hypersetup{
    colorlinks,%
    citecolor=blue,%
    filecolor=green,%
    linkcolor=red,%
    urlcolor=green
}

\begin{document}

\title{Breakdown of Kohn Theorem Near Feshbach Resonance}

\author{Hamid Al-Jibbouri}
\email[]{hamidj@fu-berlin.de}
\affiliation{Institut f\"ur Theoretische Physik, Freie Universit\"at Berlin,
Arnimallee 14, 14195 Berlin, Germany}

\author{Axel Pelster}
\email[]{axel.pelster@physik.uni-kl.de}
\affiliation{Fachbereich Physik und Forschungszentrum OPTIMAS, Technische
Universit\"at Kaiserslautern, 67663 Kaiserslautern, Germany}
\affiliation{Hanse-Wissenschaftskolleg, Lehmkuhlenbusch 4, D-27733 Delmenhorst,
Germany}

\date{\today}

\begin{abstract}
We study the collective excitation frequencies of a harmonically trapped $^{85}$Rb Bose-Einstein
condensate (BEC) in the vicinity of a Feshbach resonance. To this end, we
solve the underlying Gross-Pitaevskii (GP) equation by using a
Gaussian variational approach and obtain the coupled set of ordinary
differential equations for the widths and the center of mass of the condensate.
A linearization shows that the dipole mode frequency decreases when the bias
magnetic field approaches the Feshbach resonance, so the Kohn theorem is violated.
\end{abstract}


\maketitle

\section{Introduction}

Many experiments focus on investigating collective excitations
of harmonically trapped BECs
as they can be measured very accurately and, therefore, allow for extracting the
respective system parameters \cite{StamperKurn}. Several studies show that the
excitation of low-lying
collective modes can be achieved by modulating a system parameter. One example
is to change periodically the external potential trap
\cite{DSJin,MOMewes,YCastinandRDum,FDalfovo,JJGarciaRipoll,JJGarciaRipoll1,
AINicolin} or, more specifically, the trap anisotropy of the confining potential
\cite{FDalfovo,GHechenblaikner1,hodby,YuZhou,Hamid1,Hamid2}. Alternatively, this
can also be achieved by a periodic modulation of the s-wave scattering length
\cite{VSBagnato0,SEPollackandRGHulet,VSBagnato,IVidanovic,SSabari,AINicolin1,will} or, possibly, by modifying the
three-body interaction strength \cite{SSabari,Hamid1,Hamid2}.

In 1961 W. Kohn \cite{kohn} showed in a three-dimensional solid that the Coulomb
interaction between electrons does not change the cyclotron resonance frequency. This 
Kohn theorem can also be transfereed to the realm of ultracold quantum gases, where it states that the center of
mass of the entire cloud oscillates back and forth in the harmonic trapping
potential with the natural frequency of the trap irrespective of both the strength
and type of the two-body interaction.
The Kohn theorem for a Bose gas is discussed explicitly in the
Bogoliubov approximation at zero temperature of Ref.~\cite{DanielRokhsar}.
The dynamics of a trapped Bose condensed gas at finite temperature are consistent with a generalized Kohn theorem and
satisfy the linearized ZNG hydrodynamic equations in Ref.~\cite{Zaremba2771999}. In particular, the Kohn mode was
studied in an approximate variational approach to the kinetic theory in the
collisionless regime
in Ref.~\cite{STOOFkohn}. The validity of the Kohn theorem at finite temperature
was also shown within a linear response treatment in Ref.~\cite{Minguzzi}. Later on it
was also examined in Ref.
\cite{Jurgen} for a specific finite-temperature approximation within the
dielectric formalism.
Furthermore, the dipole mode frequency was studied by using a sum rule
approach in Refs.~\cite{sum1,sum11,sum2,sum3,sum4}. The collective dipole oscillations
in the Bose-Fermi mixture were studied theoretically in Refs.~\cite{sum11,sum2}
and experimentally in Ref.~\cite{sum5}, while the dipole oscillation of a spin-orbit
coupled Bose-Einstein condensate confined in a harmonic trap was studied
experimentally \cite{sum3} and investigated theoretically \cite{sum3,sum4}. The dipole
oscillation was also discussed
for a general fermionic mixture by using the Boltzmann equation in Ref.~\cite{sum6}.

Apart from a periodic modulation of a system parameter
the dipole mode can also be excited by introducing an
abrupt change in the potential. The experimental achievement
\cite{StamperKurn,YANGLu} has been confirmed by Refs.
\cite{CJPethick,SStringari}, where also
the quadrupole frequency was determined
as an eigenfrequency of the hydrodynamic equations.
The coupling between the internal and the external dynamics of a Bose-Einstein condensate oscillating in
an anharmonic magnetic waveguide was studied in Ref.~\cite{HOtt}. There also
several nonlinear
effects including second and third harmonic generation of the center of mass
motion, and a nonlinear mode
mixing have been identified. In the more recent work \cite{Bagnato12}, the authors
explored a different
physical
idea by investigating the coupling between dipole and quadrupole modes in the
immediate
vicinity of a Feshbach resonance.
They started with considering a Bose-Einstein condensate in a magneto-optical
Ioffe-Pritchard trap \cite{Ioffe}
with a controlled bias field, where
the dipole mode is excited. If the bias field is close enough to a Feshbach
resonance, the oscillation of the entire cloud
through the inhomogeneous bottom of the trap causes an effective periodic
time-dependent modulation in the scattering
length, which in turn changes the Kohn mode frequency, but also excites other modes like the quadrupole or the breathing mode.

Although Ref.~\cite{Bagnato12} introduces this appealing physical notion, it
only
provides a rough quantitative study. Therefore we calculate in this paper
in detail the
collective excitation modes of a harmonically trapped Bose-Einstein
condensate in the vicinity of a Feshbach resonance for experimentally realistic
parameters of a $^{85}$Rb BEC \cite{RB85,RB85Grimm}. To this end, we consider
the
situation that a
Bose-Einstein condensate oscillates within a
dipole mode in $z$-direction and investigate how the dipole mode frequency
changes when the bias magnetic
field approaches the Feshbach resonance in Section \ref{sec:NearFR}. Afterwards, we follow Ref.~\cite{Bagnato12} and
transform  the
partial differential of GP equation \cite{Grosseq,Pitaevskiieq} for the
condensate wave
function in
Section~\ref{sec:method} within a
variational approach \cite{Perezandzoller,VMPerez} into a set
of ordinary differential equations for the widths and
the center of mass position of the condensate in an axially-symmetric harmonic trap plus a bias
potential. Our analysis is based on an exact treatment with the help of the Schwinger trick
\cite{schwinger}. The resulting theory on how to determine the low-lying collective excitation frequencies
is developed step by step in Section~\ref{sec:right}. Afterwards, Section~\ref{subsec:results}  compares our results with the corresponding findings of Ref.
\cite{Bagnato12}. In addition we discuss two special cases, when the bias magnetic field field approaches the Feshbach resonance and when it is far away from the Feshbach resonance.
It turns out that the heuristic approximation in Ref.~\cite{Bagnato12}
is not valid neither on top of the Feshbach resonance, nor far away from it. Finally, in Sec.~\ref{sec:CONCLUSIONSS} we
summarize our
findings and present the conclusions.
\section{Near Feshbach Resonance}\label{sec:NearFR}
The dynamics of a
condensed Bose gas in a trap at zero temperature is described
by the time-dependent GP equation \cite{Perezandzoller,VMPerez}
\begin{equation}
i \hbar \frac{\partial \psi({\bf r},t)}{\partial t}=\Bigg[-\frac{\hbar^2}{2M}
\Delta+V_{\rm{ext.}}({\bf r}) +g_2 N n_c({\bf r},t)\Bigg] \psi({\bf r},t).
\label{eq:GP}
\end{equation}
where $\psi({\bf r},t)$ denotes a condensate wave function and $N$ represents to the total number of atoms in the condensate. On the right-hand side
of the above equation
we have a kinetic energy term, where $M$ denotes the mass
of the corresponding atomic species, an external trap $V_{\rm{ext.}}({\bf r})$, and
the third is the two-body interaction with the condensate density $n_c({\bf
r},t)=\left|\psi({\bf r},t)
\right|^{2}$ and the strength $g_2=4
\pi \hbar^2  a_{\rm s}/M$, which is proportional to
the
$s$-wave scattering length $a_{\rm s}$. In the presence of a
magnetic field, the s-wave scattering length can be tuned by applying an external
magnetic field due to the Feshbach resonance \cite{CJPethick,Moerdijk}
\begin{equation}
a_s(B)=a_{\rm{BG}}\bigg(1-\frac{\Delta}{B-B_{\rm{res}}}\bigg)\,,
\label{eq:ch1fesh}
\end{equation}
with the background s-wave scattering $a_{\rm{BG}}$, the width of the Feshbach
resonance $\Delta$, and the
resonance of magnetic field $B_{\rm{res}}$.
In this
paper, we consider a Bose-Einstein condensate confined in a magneto-optical Ioffe-Pritchard trap
composed of a cylindrically symmetric harmonic potential with trap anisotropy $\lambda$ plus a bias \cite{Bagnato12,Ioffe}:
\begin{equation}
V_{\rm{ext.}}({\bf r})=V_0+\frac{M\omega^2_{\rho}}{2} \left(\rho^2+ \lambda^2
z^2 \right) . \label{eq:externaltrap1}
\end{equation}
Due to the atomic magnetic moment $\mu_{\rm B}$ the potential is generated by a corresponding magnetic field
whose modulus is given by
\begin{equation}
B=B_0+ \frac{M \omega^2_{\rho}}{2 \mu_{\rm B}}\left(\rho^2+ \lambda^2 z^2
\right)\,,
\label{eq:magnetic}
\end{equation}
where $B_0=V_0/\mu_{\rm B}$ is
the bias field.

From Eqs.~(\ref{eq:ch1fesh}) and (\ref{eq:magnetic}), the
inter-particle interaction in the atomic cloud moving in
this potential is controlled by the spatially dependent scattering
length
\begin{equation}
a_{\rm s}=a_{\rm{BG}} \bigg[1-\frac{\Delta}{\mathcal{H} + \frac{M
\omega^2_{\rho}}{2 \mu_{\rm B}}\left(\rho^2+ \lambda^2 z^2 \right)  } \bigg]\, ,
\label{eq:controlled}
\end{equation}
where $\mathcal{H}=B_0-B_{\rm{res}}$ denotes the deviation of the bias magnetic
field $B_0$ from the location of the Feshbach resonance at $B_{\rm res}$. In the following, we consider the potential (\ref{eq:externaltrap1}) loaded with a
condensed
cloud whose dipole mode is excited in the $z$-direction. In this
configuration, far away from the Feshbach resonance the center of mass oscillates periodically at the
bottom of the trap with the Kohn mode frequency $\omega_z=\lambda
\omega_{\rho}$.

As an initial physical motivation we discuss the consequences of the Thomas-Fermi (TF) approximation. As
we assume to have a strong two-body interaction, we neglect
the kinetic energy term in
the time-independent counterpart of Eq.~(\ref{eq:GP}) and obtain
\begin{equation}
\mu=V_{\rm{ext.}}({\bf r}) +g_2 n_c({\bf r}).
\label{eq:time-indepGPch6}
\end{equation}
Far away from the Feshbach resonance we can consider the potential contribution in
Eq.~(\ref{eq:controlled}) to be small, thus we expand
Eq.~(\ref{eq:controlled}) up to the first order of the external potential,
yielding
\begin{align}
\mu=V_{\rm{ext.}}({\bf r}) &+\frac{4 \pi \hbar^2 a_{\rm{BG}} n_c({\bf 0}) }{M}
\bigg[1-\frac{\Delta}{\mathcal{H}}\nonumber\\&+\frac{\Delta M\omega^2_{\rho}}{2\mathcal{H}^2\mu_{\rm B}} \left(\rho^2+ \lambda^2
z^2
\right)  +.....\bigg]\,,
\label{eq:time-indepGPch61}
\end{align}
where $n_c({\bf 0})$ is the TF density at the trap center with the chemical
potential $\mu=\frac{\hbar \omega_{\rho}}{2}\left(\frac{15N \lambda a_{\rm eff}}{l}\right)^{2/5}$.
On the one hand we read off from Eq.~(\ref{eq:time-indepGPch61}) an effective
s-wave scattering length
\begin{equation}
 a_{\rm eff}=a_{\rm BG} \left( 1-\frac{\Delta}{\mathcal{H}}\right)
 \label{eq:scatteringeffective}\,.
\end{equation}
In the following discussion we have a ${}^{85}$Rb BEC in mind, whose Feshbach resonance is
characterized by a negative background value of the s-wave scattering length, i.e.~$a_{\rm{BG}}<0$, and a positive width, i.e.~$\Delta >0$ \cite{RB85,RB85Grimm}.
Thus, the BEC is unstable, i.e.~$a_{\rm eff}<0$, provided that $B_0<B_{\rm
crit}+\Delta$. Conversely, the TF approximation yields a stable BEC, i.e.~$a_{\rm eff}>0$,
in the case that $B_{\rm res} <B_0<B_{\rm crit}=B_{\rm res}+\Delta$. On the
other hand, we obtain from Eqs.~(\ref{eq:externaltrap1}) and
(\ref{eq:time-indepGPch61}) an effective Kohn mode frequency
\begin{equation}
 \omega_{D,\rm{eff}}=\lambda \omega_{\rho} \sqrt{ 1+ \frac{4 \pi \hbar^2
a_{\rm{BG}} n_c({\bf 0}) \mu \Delta }{M \mathcal{H}^2\mu_{\rm
B}}}\label{eq:omegaeffective}\,.
\end{equation}
Thus, on the right-hand side of the Feshbach resonance, i.e.~for $B_{\rm res}
<B_0<B_{\rm crit}=B_{\rm res}+\Delta$, we expect due to $a_{\rm BG} <0$ that the Kohn mode frequency
Eq.~(\ref{eq:omegaeffective}) is smaller than
the corresponding one without the Feshbach resonance. In the following we will
show that this initial qualitative finding is confirmed by a more quantitative
analysis. In particular, it will turn out that the leading change of the Kohn mode frequency far away from the Feshbach resonance is, indeed, of the order $1/\mathcal{H}^2$.
\section{Variational Approach}\label{sec:method}
Equation (\ref{eq:GP}) can be cast into a variational problem, which corresponds
to the extremization of the action defined by the Lagrangian
\begin{eqnarray}
L(t)&=&\int d{\bf r} \bigg[\frac{i \hbar}{2}\left(\psi^* \frac{\partial
\psi}{\partial
t}- \psi\frac{\partial \psi^*}{\partial t} \right)\nonumber\\&&
\hspace{-0.35cm}- \ \ \frac{\hbar^2}{2M}|\nabla
\psi|^2-V({\bf r})|\psi|^2 -  \frac{g_2N}{2}
|\psi|^4\, \bigg].
\label{eq:lagrangedensity}
\end{eqnarray}
In order to analytically study the dynamical system of a BEC with two-body
contact interaction, where the dipole mode is excited in $z$-direction, we
use a Gaussian variational ansatz which includes the center of mass oscillation in
the $z$-direction according to Refs.~\cite{Perezandzoller,VMPerez,Bagnato12}.
For an axially symmetric trap, this time-dependent
ansatz reads
\begin{eqnarray}
\hspace{-0.5cm}\psi^G(\rho,z,t)&=&{\mathcal N}(t)
\exp\bigg[-\frac{\rho^2}{2u_{\rho}^2}+i
\rho \alpha_{\rho}+i
\rho^2 \beta_{\rho}\bigg]  \nonumber\\&& \hspace{-0.3cm}\times
\exp\bigg[-\frac{(z-z_0)^2}{2u_z^2}+i
z \alpha_z+i z^2 \beta_z\bigg],
\label{eq:Gch5}
\end{eqnarray}
where $\mathcal{N}=1/\sqrt{\pi^{\frac{3}{2}} u_{\rho}^2 u_z}$ is a normalization
factor, while $u_{\rho,z}$, $z_0$, $\alpha_{\rho,z}$, and $\beta_{\rho,z}$
denote time-dependent variational parameters, which represent
radial and axial condensate widths, the center of mass position, and the
corresponding phases.
Inserting the Gaussian ansatz (\ref{eq:Gch5}) into the
Lagrange function~(\ref{eq:lagrangedensity}), we
obtain
\begin{eqnarray}
&& L(t)=- \frac{\hbar^2}{2M} \bigg[\frac{1}{2 u_{z}^2}+\frac{1}{u_{\rho}^2}+2
u_{z}^2
\beta_{z}^2+4 z_0^2 \beta_{z}^2+4 z_0 \beta_{z}
\alpha_{z}\nonumber\\&&+\alpha_{z}^2+4 u_{\rho}^2
\beta_{\rho}^2+2 \sqrt{\pi} u_{\rho} \beta_{\rho}
\alpha_{\rho} +\alpha_{\rho} ^2 \bigg]-\frac{\hbar}{2} \bigg[u_{z}^2
\dot{\beta}_{z}+2 z_{0}^2
\dot{\beta}_{z}\nonumber\\&&+2 z_{0} \dot{\alpha}_{z}+2 u_{\rho}^2
\dot{\beta}_{\rho}+\sqrt{\pi} u_{\rho}
\dot{\alpha}_{\rho} \bigg] -V_0-\frac{\hbar^2 N
a_{\rm{BG}}}{\sqrt{2 \pi} M} \frac{1}{u^2_{\rho} u_z} \nonumber\\&&-
\frac{M \omega^2_{\rho}}{2} \bigg[ u^2_{\rho}+\frac{\lambda ^2 u^2_{z}}{2}+
\lambda^2 z_0^2\bigg] + \frac{4 \hbar^2 N
a_{\rm{BG}}\Delta}{\pi u^4_{\rho} u_z^2M} f\, ,
\label{eq:Lagrangian function}
\end{eqnarray}
where we have introduced the integral
\begin{equation}
 \hspace{-0.07cm}f=\int_0^{\infty}  \hspace{-0.3cm}d\rho \int_{-\infty}^{\infty}
 \hspace{-0.3cm}dz
\frac{\rho
\exp\bigg[-2\rho^2/u_{\rho}^2-2(z-z_0)^2/u_z^2
\bigg]}{\mathcal{H} + \frac{M \omega^2_{\rho}}{2 \mu_{\rm B}}\left(\rho^2+
\lambda^2
z^2 \right)}.
\label{eq:fintegral}
\end{equation}
From the corresponding Euler-Lagrange equations we
obtain the equations of motion for all
variational parameters. The phases $\alpha_{\rho,z}$ and $\beta_{\rho,z}$ can be
expressed explicitly in terms of first derivatives of the widths $u_{\rho}$,
$u_z$, and the center of mass coordinate $z_0$ according to
\begin{align}
\alpha_{\rho}&=0 \,, \hspace{0.35cm} \alpha_{z}=\frac{M}{\hbar}\dot{z}_0-2 z_0
\beta_z \,, \hspace{0.35cm}
\beta_{\rho,z}=\frac{M}{2 \hbar}\frac{\dot u_{\rho,z}}{u_{\rho,z}}\,.
\label{eq:phases}
\end{align}
Inserting Eq.~(\ref{eq:phases}) into the Euler-Lagrange equations for the width
of the condensates $u_{\rho}$, $u_z$, and the center of mass coordinate $z_0$,
we obtain a system of three
second-order differential equations for $u_{\rho}$, $u_z$, and $z_0$:
After rescaling the quantities according to
\begin{eqnarray}
u_i,\rho,z,z_0 \rightarrow l (u_i,\rho,z,z_0) , \ \
t \rightarrow t \omega_{\rho}\,,
\end{eqnarray}
with the oscillating length $l=\sqrt{\hbar/(M \omega_{\rho})}$, we obtain a system
of second-order differential equations for $u_{\rho}$,
$u_z$, and $z_0$ in the dimensionless form \cite{Bagnato12}
 \begin{align}
 &\ddot{u}_{\rho}+ u_{\rho}-\frac{1}{u_{\rho}^3}
-\frac{\mathcal{P}_{\rm{BG}}}{u_{z}
u_{\rho}^3}\label{eq:urf}\\& \hspace{0.5cm}\times \bigg[1-\frac{16 \varepsilon_0
f}{\sqrt{2\pi} l^3 u_{\rho}^2 u_z}+
\frac{4
\varepsilon_0}{\sqrt{2\pi} l^2 u_{\rho} u_z} \frac{\partial f}{\partial
u_{\rho}}
\bigg]=0 \, ,\nonumber\\
 &\ddot{u}_{z}+ \lambda^2 u_{z}- \frac{1}{u_{z}^3}-\frac{\mathcal{P}_{\rm{BG}}
}{u_{z}^2
u_{\rho}^2}\label{eq:uzf}\\& \hspace{0.5cm}\times\bigg[1-\frac{16 \varepsilon_0
f}{\sqrt{2 \pi} l^3 u_{\rho}^2 u_z}+
\frac{8
\varepsilon_0}{\sqrt{2 \pi} l^2 u_{\rho}^2}  \frac{\partial f}{ \partial u_z}
\bigg]=0 \,, \nonumber\\
&\ddot{z}_{0} + \lambda^2 z_{0}-\frac{4
\mathcal{P}_{\rm{BG}}\varepsilon_0}{\sqrt{2\pi} l^2 u^4_{\rho} u_z^2}
\frac{\partial
f}{\partial z_0}=0\label{eq:z0f}\,.
\end{align}
Here we have introduced the dimensionless parameters
\begin{eqnarray}
\mathcal{P}_{\rm{BG}}=\sqrt{\frac{2}{\pi}} \frac{N a_{\rm{BG}}}{l}, \,
\varepsilon_0=\frac{\Delta}{\mathcal{H}}, \,
\varepsilon_1=\frac{\mathcal{H} \mu_{\rm B}}{(\hbar \omega_{\rho})}, \,
\varepsilon=\varepsilon_0 \,\varepsilon_1 \,.
\label{eq:dimensionlessparameter}
\end{eqnarray}
In order to study the frequencies of collective modes both in the vicinity of
the Feshbach resonance and on the right-hand side of the Feshbach resonance, i.e.~for $\mathcal{H}>0$, we develop now our own approach by using the Schwinger
trick \cite{schwinger} in order to rewrite the integral Eq.~(\ref{eq:fintegral})
in form of
\begin{align}
f&=l^3 \int_0^{\infty} \hspace{-0.3cm}d\rho \int_{-\infty}^{\infty}
\hspace{-0.3cm}dz
\int_0^{\infty}\hspace{-0.3cm} d\mathcal{S}  \  \rho
\exp\bigg[-\frac{2\rho^2}{u_{\rho}^2}-\frac{2(z-z_0)^2}{u_z^2}
\bigg] \nonumber\\& \hspace{0.4cm}\times \exp \bigg[-\mathcal{S}-\frac{\mathcal{S}}{2
\varepsilon_1}\left(\rho^2+
\lambda^2 z^2 \right) \bigg] \, . \label{eq:fintegral1}
\end{align}
In the following, we concentrate on the topic how
this violates the Kohn theorem,
i.e.~how the dipole mode frequency changes when the bias magnetic field $B_0$
approaches the Feshbach resonance $B_{\rm res}$. Within the linearization of the
equations of motions
(\ref{eq:urf})--(\ref{eq:z0f}), we have to take into the account that the
equilibrium value of the center of mass
position vanishes according to Eq.~(\ref{eq:z0f}). This allows to expand the
integral of Eq.~(\ref{eq:fintegral1}) up to
the second order of $z_0$, which yields
\begin{eqnarray}
f&\hspace{-0.1cm}=&\hspace{-0.1cm} l^3 \hspace{-0.1cm}\int_0^{\infty} \hspace{-0.3cm}d\rho
\int_{-\infty}^{\infty}
\hspace{-0.3cm}dz
\int_0^{\infty}\hspace{-0.3cm} d\mathcal{S} \ \rho \bigg[1+\frac{4 z
z_0}{u_z^2}-\frac{2z_0^2}{u_z^2}+\frac{8 z^2z_0^2}{u_z^4}+.. \bigg]
\nonumber\\&& \times\exp
\left[-\frac{2 \rho^2}{u_{\rho}^2}-\frac{2
z^2}{u_z^2}-\mathcal{S}-\frac{\mathcal{S}}{2\varepsilon_1}\left(\rho^2+
\lambda^2 z^2\right)   \right]\,.
\label{eq:expandffunction}
\end{eqnarray}
Correspondingly we determine the respective first derivatives $\frac{\partial f}{\partial
u_{\rho}}$, $\frac{\partial f}{\partial u_z}$, and $\frac{\partial f}{\partial
z_0}$ which appear in the equations of motion (\ref{eq:urf})--(\ref{eq:z0f}).
\section{Right-Hand Side of Feshbach Resonance}\label{sec:right}
We consider in this section the frequencies of
collective modes when the bias field $B_0$ is larger than or equal to the
resonant magnetic field $B_{\rm{res}}$, i.e.~$\mathcal{H}=B_0-B_{\rm{res}}\geq0$.
\subsection{Collective Mode Frequencies}\label{subsec:collectivemodefrequencies}
At first we obtain a system of three second-order ordinary differential equations
for
$u_{\rho}$, $u_z$, and $z_0$ in the dimensionless form after inserting
Eq.~(\ref{eq:expandffunction}) into Eqs.~(\ref{eq:urf})--(\ref{eq:z0f}):
\begin{eqnarray}
\ddot{u}_{\rho}+&& \hspace{-0.4cm} u_{\rho}-\frac{1}{u_{\rho}^3}-
\frac{\mathcal{P}_{\rm{BG}}
}{u_{z}
u_{\rho}^3}\label{eq:urB0}\\&&\hspace{-0.7cm}\times\bigg[1 -16 \hspace{-0.1cm}
\int_0^{\infty}\hspace{-0.2cm}
\frac{\varepsilon \
\varepsilon_1^{1/2}
d\mathcal{S}\ e^{-\mathcal{S}}
\left(2\varepsilon_1+\mathcal{S} u_{\rho}^2 \right)}{\left(4
\varepsilon_1+\mathcal{S} u_{\rho}^2 \right)^2 \sqrt{4\varepsilon_1 +\mathcal{S}
u_{z}^2 \lambda ^2 }}+...
\bigg]=0,\nonumber\\
\ddot{u}_{z}+ && \hspace{-0.4cm} \lambda^2 u_{z}-
\frac{1}{u_{z}^3} -\frac{\mathcal{P}_{\rm{BG}} }{u_{z}^2
u_{\rho}^2}\label{eq:uzB0}\\&& \hspace{-0.7cm}\times \bigg[1-16
 \hspace{-0.1cm}
\int_0^{\infty}\hspace{-0.2cm}
\frac{\varepsilon \ \varepsilon_1^{1/2} d\mathcal{S}\ e^{-\mathcal{S}}
\left(2 \varepsilon_1 + \mathcal{S} u_{z}^2 \lambda ^2 \right)}{\left(4
\varepsilon_1 + \mathcal{S} u_{\rho}^2
\right)
\left(4 \varepsilon_1  +\mathcal{S} u_{z}^2 \lambda ^2 \right)^{3/2}}+...
\bigg]=0,\nonumber\\
\ddot{z}_{0} + &&\hspace{-0.4cm}\lambda^2 z_0  \bigg[1+\frac{16  \mathcal{P}_{\rm{BG}}
}{u^2_{\rho} u_z} \label{eq:z0B0}\\&&
\times \int_0^{\infty}
\hspace{-0.2cm}\frac{\varepsilon \ \varepsilon_1^{1/2} d\mathcal{S} \
e^{-\mathcal{S}} \mathcal{S}}{\left(4 \varepsilon_1 + \mathcal{S} u_{\rho}^2
\right)
\left(4
\varepsilon_1+\mathcal{S} u_z^2 \lambda ^2 \right)^{3/2}}+...\bigg]=0\, \nonumber\,.
\end{eqnarray}
The time-independent solution of the condensate widths $u_{\rho}=u_{\rho0}$,
$u_z=u_{z0}$, and $z_0=z_{00}$ is determined from
\begin{eqnarray}
&&\hspace{-1cm}u_{\rho0}-\frac{1}{u_{\rho0}^3}- \frac{\mathcal{P}_{\rm{BG}}
}{u_{z0}
u_{\rho0}^3}\bigg[1 -16 \varepsilon \ \varepsilon_1^{1/2} \nonumber\\&&
\hspace{-0.3cm} \times
\int_0^{\infty}\hspace{-0.3cm}
\frac{
d\mathcal{S}\ e^{-\mathcal{S}}
\left(2\varepsilon_1+\mathcal{S} u_{\rho0}^2 \right)}{\left(4
\varepsilon_1+\mathcal{S} u_{\rho0}^2 \right)^2 \sqrt{4\varepsilon_1
+\mathcal{S}
u_{z0}^2 \lambda ^2 }}
\bigg]=0\,,\label{eq:ur0B0}
\end{eqnarray}
\begin{eqnarray}
&&\hspace{-1cm} \lambda^2 u_{z0}-
\frac{1}{u_{z0}^3} -\frac{\mathcal{P}_{\rm BG} }{u_{z0}^2 u_{\rho0}^2}\bigg[1-16
\varepsilon \ \varepsilon_1^{1/2}  \nonumber\\&& \hspace{-0.3cm} \times
\int_0^{\infty}\hspace{-0.3cm}
\frac{d\mathcal{S}\ e^{-\mathcal{S}}
\left(2 \varepsilon_1 + \mathcal{S} u_{z0}^2 \lambda ^2 \right)}{\left(4
\varepsilon_1 + \mathcal{S} u_{\rho0}^2
\right)
\left(4 \varepsilon_1  +\mathcal{S} u_{z0}^2 \lambda ^2 \right)^{3/2}}
\bigg]
=0,
\label{eq:uz0B0}\\
&&\hspace{-1cm}z_{00}=0\, \label{eq:z00B0} \, .
\end{eqnarray}
Using the Gaussian approximation enables us to analytically estimate
the frequencies of the low-lying collective modes
\cite{Perezandzoller,VMPerez,IVidanovic,Hamid1,Hamid2} and the dipole mode
frequency. This
is done by linearizing Eqs.~(\ref{eq:urB0})--(\ref{eq:z0B0}) around the
equilibrium
positions Eqs.~(\ref{eq:ur0B0})--(\ref{eq:z00B0}). If we expand the condensate widths as $u_{\rho}=u_{\rho 0}+\delta
u_{\rho}$, $u_z=u_{z0}+\delta u_z$, and the center of mass motion as
$z_0=z_{00}+\delta z_0$, insert these expressions into
the corresponding equations, and expand them around the equilibrium widths by
keeping only linear terms, we immediately get for the breathing and quadrupole
frequencies
\begin{align}
\omega_{B,Q}^2=\frac{1}{2}\bigg[m_1+m_3\pm
\sqrt{(m_1-m_3)^2+8m_2^2}\,\bigg]\,,
\label{eq:frequencytop}
\end{align}
where the abbreviations
$m_1, m_2$ and $m_3$ are calculated by using
Mathematica \cite{Mathematica}:
\begin{eqnarray}
m_1=1+\frac{3}{u_{\rho0}^4}&&\hspace{-0.4cm}+\frac{3 \mathcal{P}_{\rm
BG}}{u_{\rho0}^4
u_{z0}} \bigg[1-16 \varepsilon \ \varepsilon_1^{1/2} \hspace{-0.15cm}
\label{eq:m1} \\&& \hspace{-2cm} \times \int_0^{\infty} \frac{d\mathcal{S}
e^{-\mathcal{S}} \left(5 \mathcal{S}^2
u_{\rho0}^4+18 \mathcal{S} u_{\rho0}^2 \varepsilon_1+24
\varepsilon_1^2\right)}{3
\left(\mathcal{S} u_{\rho0}^2+4 \varepsilon_1\right)^3 \sqrt{4
\varepsilon_1+\mathcal{S} u_{z0}^2 \lambda ^2}} \bigg]\,  ,\nonumber\\
&& \hspace{-2.5cm} m_2= \frac{\mathcal{P}_{\rm BG}}{u_{\rho0}^3 u_{z0}^2}
\bigg[1-32 \varepsilon \
\varepsilon_1^{1/2} \hspace{-0.15cm} \label{eq:m2} \\&& \hspace{-2cm} \times
\int_0^{\infty}\frac{d\mathcal{S} e^{-\mathcal{S}}
\left(\mathcal{S} u_{\rho0}^2+2 \varepsilon_1\right) \left(2
\varepsilon_1+\mathcal{S} u_{z0}^2 \lambda ^2\right)}{\left(\mathcal{S}
u_{\rho0}^2+4 \varepsilon_1\right)^2
\left(4 \varepsilon_1+\mathcal{S} u_{z0}^2 \lambda ^2\right)^{3/2}} \bigg]\, ,
\nonumber\\
m_3=\lambda ^2+\frac{3}{u_{z0}^4}&&\hspace{-0.4cm}+\frac{2 \mathcal{P}_{\rm
BG}}{u_{\rho0}^2
u_{z0}^3} \bigg[1 -8 \varepsilon \ \varepsilon_1^{1/2} \hspace{-0.15cm}
\label{eq:m3}\\&& \hspace{-2cm} \times \int_0^{\infty}
 \frac{d\mathcal{S} e^{-\mathcal{S}} \left(16 \varepsilon_1^2+10 \mathcal{S}
u_{z0}^2 \varepsilon_1
 \lambda ^2+3 \mathcal{S}^2 u_{z0}^4 \lambda ^4\right)}{\left(\mathcal{S}
u_{\rho0}^2+
 4 \varepsilon_1\right) \left(4 \varepsilon_1+\mathcal{S} u_{z0}^2 \lambda
^2\right)^{5/2}} \bigg]\,\nonumber .
\end{eqnarray}
The quadrupole mode has a lower frequency and is characterized by out-of phase
radial and axial oscillations, while in-phase oscillations correspond to the
breathing
mode. Furthermore, the dipole mode frequency is given by
\begin{eqnarray}
\omega_D^2=&&\hspace{-0.4cm}\lambda^2 \bigg[1+
\frac{16\mathcal{P}_{\rm{BG}} }{u^2_{\rho0} u_{z0}}
 \int_0^{\infty}\hspace{-0.35cm}
\frac{\varepsilon \
\varepsilon_1^{1/2} d\mathcal{S} \
e^{-\mathcal{S}} \mathcal{S}}{\left(4 \varepsilon_1 + \mathcal{S} u_{\rho0}^2
\right) \left(4 \varepsilon_1+\mathcal{S} u_{z0}^2 \lambda ^2 \right)^{3/2}}
\bigg]\label{eq:dipoleB0} \,.\nonumber\\
\end{eqnarray}
\subsection{Thomas-Fermi Approximation}\label{subsec:thomasfermiapproximation}
In order to find an analytical description for
the condensate widths $u_{\rho0}$, $u_{z0}$, and their ratio $u_{\rho0}/u_{z0}$
as well as the frequencies of collective modes, we consider now the TF approximation. Thus, we neglect the respective second term in
Eqs.~(\ref{eq:ur0B0}), (\ref{eq:uz0B0}),
which comes from the kinetic energy. Furthermore, we use the ansatz
\begin{equation}
\frac{u_{z0} \lambda}{u_{\rho0}}=1+\eta
\label{eq:eta}
\end{equation}
and evaluate the resulting equations in the limit of a vanishing smallness parameter $\eta$, yielding
\begin{align}
u_{\rho 0}&- \frac{\mathcal{P}_{\rm{BG}} }{u_{z 0}
u_{\rho 0}^3}\bigg[1 -16 \varepsilon \ \varepsilon_1^{1/2}
\hspace{-0.2cm}\int_0^{\infty}\hspace{-0.3cm}
 d\mathcal{S} \frac{e^{-\mathcal{S}} \left(\mathcal{S} u_{\rho 0}^2+2
\varepsilon_1\right)}{\left(\mathcal{S} u_{\rho 0}^2+4
\varepsilon_1\right)^{5/2}} \bigg]=0 \label{eq:ur0etaB0}\, ,\\
\lambda^2 u_{z 0}&-\frac{\mathcal{P}_{\rm{BG}} }{u_{z 0}^2 u_{\rho
0}^2}\bigg[1-16
\varepsilon \ \varepsilon_1^{1/2} \hspace{-0.2cm} \int_0^{\infty}\hspace{-0.3cm}
d\mathcal{S} \frac{e^{-\mathcal{S}}\left(\mathcal{S} u_{\rho 0}^2+2
\varepsilon_1\right)}{\left(\mathcal{S} u_{\rho 0}^2+4
\varepsilon_1\right)^{5/2}}\bigg]=0 \label{eq:uz0etaB0}\, .
\end{align}
Solving the remaining $\mathcal{S}$-integral we obtain the equilibrium
widths $u_{\rho0}$ and $u_{z0}$ in TF approximation
\begin{eqnarray}
 &&u_{\rho 0}^5- \mathcal{P}_{\rm BG} \lambda \label{eq:ur0TF}\\&&
\hspace{0.3cm}\times\bigg[1-\frac{\varepsilon}{3}\bigg(\frac{40
}{u_{\rho0}^2}+\frac{64 \varepsilon_1}{u_{\rho0}^4}\bigg)+ \left(3 u_{\rho0}^2+
4 \varepsilon_1 \right)\kappa
\bigg]=0\, ,\nonumber\\
&&\lambda u_{z 0}=u_{\rho0}\,,
\label{eq:uz0TF}
\end{eqnarray}
where we have introduced the abbreviation
\begin{equation}
 \kappa=\frac{8\varepsilon\sqrt{\pi \varepsilon_1}}{u_{\rho0}^5}
e^{4\varepsilon_1/u_{\rho0}^2}\text{Erfc}\left[2\sqrt{\varepsilon_1}/u_{\rho0
}\right]\,,
\label{eq:kapp}
\end{equation}
with the complementary error function:
\begin{equation}
 \text{Erfc}(x)=\frac{2}{\sqrt{\pi}} \int_x^{\infty} dt \ e^{-t^2}\,.
 \label{eq:ERFC}
\end{equation}
In the
similar way we obtain the quadrupole, breathing, and dipole mode frequencies in
TF approximation by inserting
Eq.~(\ref{eq:eta}) into Eqs.~(\ref{eq:m1})--(\ref{eq:dipoleB0}) and evaluating the limit $\eta\rightarrow 0$. Solving
the remaining $\mathcal{S}$-integrals we obtain analytically the quadrupole and breathing
frequencies in
TF approximation via Eq.~(\ref{eq:frequencytop}) with the abbreviations
\begin{eqnarray}
m_1&=&1+\frac{3 \mathcal{P}_{\rm{BG}} \lambda }{u_{\rho0}^5}\bigg[1-\frac{8
\varepsilon }{45 u_{\rho0}^{7}} \bigg( 107 u_{\rho0}^5+408
u_{\rho0}^3 \varepsilon_1\nonumber\\&&\hspace{-1cm}+256 u_{\rho0}
\varepsilon_1^2\bigg) +\frac{\kappa }{45}  \bigg(300 u_{\rho0}^2+880
\varepsilon_1 +\frac{512
 \varepsilon_1^2}{u_{\rho0}^2}\bigg)\bigg]\,,\label{eq:m1etaB0}\\
 m_2&=&\frac{\mathcal{P}_{\rm BG} \lambda ^2}{u_{\rho 0}^5}\bigg[1-\frac{8
\varepsilon }{15 u_{\rho0}^{7}}\bigg(43 u_{\rho0}^5+152 u_{\rho0}^3
\varepsilon_1\nonumber\\&&\hspace{-1cm}+64 u_{\rho0}
\varepsilon_1^2\bigg)+\frac{\kappa }{15 }
\bigg(120 u_{\rho0}^2+320 \varepsilon_1+\frac{128 \varepsilon_1^2}{u_{\rho0}^2}
\bigg)\bigg]\,,\label{eq:m2etaB0}\\
 m_3&=&\lambda ^2+\frac{2 \mathcal{P}_{\rm{BG}} \lambda ^3}{u_{\rho
0}^5}\bigg[1-\frac{16 \varepsilon}{15 u_{\rho0}^{7}} \bigg( 16 u_{\rho0}^5
+ 64u_{\rho0}^3 \varepsilon_1\nonumber\\&&
\hspace{-1cm}+48u_{\rho0}\varepsilon_1^2\bigg) +\frac{\kappa}{15} \bigg(90
u_{\rho0}^2+280 \varepsilon_1+\frac{192
\varepsilon_1^2}{u_{\rho0}^2} \bigg) \bigg]\label{eq:m3etaB0}\,,
\end{eqnarray}
whereas the dipole mode frequency in TF approximation reads explicitly
\begin{align}
\omega_D^2&= \lambda^2+\frac{32 \mathcal{P}_{\rm{BG}} \lambda^3 \varepsilon }{3
u_{\rho0}^{10}}   \bigg[u_{\rho0}^3+4
u_{\rho0} \varepsilon_1-\frac{\kappa u_{\rho0}^5}{8 \varepsilon} \nonumber\\& \hspace{3cm} \hspace{-2.5cm}\times \bigg(3
u_{\rho0}^2+8 \varepsilon_1\bigg)\bigg] \label{eq:dipoleetaB0}\,.
\end{align}
%
\subsection{On Top of Feshbach Resonance}\label{subsec:ontopoffeshbachresonance}
Now, as a physically important special case, we apply the TF
approximation to the condensate widths Eqs.~(\ref{eq:ur0TF}), (\ref{eq:uz0TF})
and to the frequencies of collective modes Eq.~(\ref{eq:frequencytop})
where the abbreviations $m_1$, $m_2$, and $m_3$ are
defined in Eqs.~(\ref{eq:m1etaB0})--(\ref{eq:dipoleetaB0}) on top of the
Feshbach resonance. In the limit $\mathcal{H}\rightarrow 0$ or
$\varepsilon_1 \rightarrow 0$ we obtain the condensate widths
 \begin{align}
u_{\rho0}^5&-\mathcal{P}_{\rm{BG}} \lambda\left(1-  \frac{40\varepsilon}{3
u_{\rho0}^2}\right)=0
 \label{eq:ur0topTF}\, ,\\
\lambda u_{z 0}&=u_{\rho0}\,,
\label{eq:uz0topTF}
\end{align}
the breathing and quadrupole frequencies (\ref{eq:frequencytop}) from
\begin{eqnarray}
&&m_1=1+\frac{3\mathcal{P}_{\rm{BG}} \lambda}{u_{\rho
0}^5} \left( 1-  \frac{856\varepsilon}{45u_{\rho0}^2} \right)\label{eq:m1eta}\,
,\\
&&m_2=\frac{\mathcal{P}_{\rm{BG}}
\lambda^2}{u_{\rho0}^5}\left(1-\frac{344\varepsilon}{15u_{\rho0}^2}\right)\label
{eq:m2eta}\, ,\\
&&m_3=\lambda^2  + \frac{2\mathcal{P}_{\rm{BG}} \lambda^3}{u_{\rho 0}^5 }\left(1
-\frac{256\varepsilon}{15u_{\rho0}^2}\right)\label{eq:m3eta}\,.
\end{eqnarray}
and the dipole mode frequency
\begin{align}
\omega^2_D&=\lambda^2+ \frac{32 \varepsilon \lambda^3 \mathcal{P}_{\rm
BG}}{3u^7_{\rho0}}
\label{eq:dipoletopeta}\,.
\end{align}
All these results on top of the Feshbach resonance turn out to be finite in contrast to the finiding of Ref.~\cite{Bagnato12}.

\subsection{Far Away from Feshbach Resonance}\label{subsec:farawayfeshbachresonance}
Accordingly, we also apply the TF approximation to the condensate widths
Eqs.~(\ref{eq:ur0etaB0}), (\ref{eq:uz0etaB0}) and to the frequencies of collective
modes Eq.~(\ref{eq:frequencytop}), where the abbreviations $m_1$, $m_2$, and $m_3$ are defined in
Eqs.~(\ref{eq:m1etaB0})--(\ref{eq:m3etaB0}), (\ref{eq:dipoleetaB0})
for the case when $B_0$ is far away from the Feshbach resonance. In the limit $\mathcal{H}\rightarrow
\infty$ or $\varepsilon_1 \rightarrow \infty$ we have to expand the complementary error function
(\ref{eq:ERFC}) for large real $x$
\begin{equation}
 \text{Erfc}(x)= \frac{e^{-x^2}}{\sqrt{\pi }} \left(\frac{1}{ x}-\frac{1}{2
x^3}+\frac{3}{4  x^5}+....\right)\,,
 \label{eq:expanderfc}
\end{equation}
yielding a corresponding asymptotic expansion for $\kappa$ from Eq.~(\ref{eq:kapp})
\begin{equation}
 \kappa=8\varepsilon \left( \frac{1}{2u_{\rho0}^4}-\frac{1}{16 u_{\rho0}^2
\varepsilon_1}+\frac{3}{128
\varepsilon_1^2}+.... \right)\,.
\label{eq:kapp1}
\end{equation}
Inserting the expansion (\ref{eq:kapp1}) into Eqs.~(\ref{eq:ur0TF}),
(\ref{eq:m1etaB0})--(\ref{eq:dipoleetaB0}), we get for the condensate widths:
\begin{align}
u_{\rho0}^5&- \mathcal{P}_{\rm{BG}} \lambda \left(1-\varepsilon_0 +\frac{
u_{\rho0}^2\varepsilon_0}{8\varepsilon_1}+....\right)=0
\label{eq:ur0TFvarepsiloninfty}\, ,\\
\lambda u_{z 0}&=u_{\rho0}\,,
\label{eq:uz0TFvarepsiloninfty}
\end{align}
the breathing and quadrupole frequencies Eq.~(\ref{eq:frequencytop}) are given by 
\begin{eqnarray}
\hspace{-0.7cm}m_1\hspace{-0.2cm}&=\hspace{-0.2cm}&1+\frac{3
\mathcal{P}_{\rm{BG}} \lambda }{u_{\rho0}^5}\bigg(1-\varepsilon_0+\frac{
\varepsilon_0 u_{\rho0}^2}{8 \varepsilon_1}-\frac{17
u_{\rho0}^4\varepsilon_0}{192
\varepsilon_1^2}+..\bigg)\label{eq:m1etaB0varepsiloninfty},\\
\hspace{-0.7cm} m_2\hspace{-0.2cm}&=\hspace{-0.2cm}&\frac{\mathcal{P}_{\rm BG}
\lambda ^2}{u_{\rho 0}^5}\bigg(1-\varepsilon_0 -\frac{\varepsilon_0 u_{\rho0}^2
}{8 \varepsilon_1}+\frac{17 u_{\rho0}^4\varepsilon_0}{64
\varepsilon_1^2}+..\bigg)\label{eq:m2etaB0varepsiloninfty},\\
\hspace{-0.7cm}m_3\hspace{-0.2cm}&=\hspace{-0.2cm}&\lambda ^2+\frac{2
\mathcal{P}_{\rm{BG}} \lambda ^3}{u_{\rho
0}^5}\bigg(1-\varepsilon_0+\frac{\varepsilon_0 u_{\rho0}^2}{4
\varepsilon_1}-\frac{17 u_{\rho0}^4\varepsilon_0}{64
\varepsilon_1^2}+..\bigg)\label{eq:m3etaB0varepsiloninfty},
\end{eqnarray}
and for the dipole frequency
\begin{align}
\omega^2_D&=\lambda^2 \left(1+\frac{\varepsilon_0
\mathcal{P}_{\rm{BG}}\lambda }{2\varepsilon_1 u_{\rho 0}^3}+... \right)
\label{eq:dipoleetaB0varepsiloninfty}\,.
\end{align}
These results for $B_0$ far away from the Feshbach resonance are now compared with the corresponding findings of Ref.~\cite{Bagnato12},
which we elaborate briefly in the next subsection.
\subsection{Heuristic Approximation}\label{subsec:heuristicapproximation}
In this section we discuss the heuristic approximation of Ref.~\cite{Bagnato12} for 
evaluating the integral (\ref{eq:fintegral}). To this end we
assume that the cloud size is much smaller than the oscillating amplitude, which
means that the cloud experiences the same
field at any point, i.e., the scattering length is homogeneous in the entire cloud. This is equivalent to stating that the numerator of
the integral (\ref{eq:fintegral}), i.e.
\begin{equation}
\rho
\exp\left[-2\rho^2/u_{\rho}^2-2(z-z_0)^2/u_z^2\right]
\label{eq:fBagnato}\, ,
\end{equation}
is much narrower than the denominator
\begin{equation}
\frac{1}{\mathcal{H} + \frac{M
\omega^2_{\rho}}{2 \mu_{\rm B}}\left(\rho^2+ \lambda^2 z^2 \right)  }\, ,
\label{eq:fBagnato1}
\end{equation}
which leads to the conditions
\begin{align}
&u_{\rho}\ll \sqrt{\frac{2 \mu_{\rm B} \mathcal{H}}{M \omega_{\rho}^2}}\,, \quad
\quad u_z\ll \sqrt{\frac{2 \mu_{\rm B} \mathcal{H}}{M
\omega_{\rho}^2\lambda^2}}\,. \label{eq:conditionch5}
\end{align}
Thus, the heuristic approximation of Ref.~\cite{Bagnato12} seems to be valid for a large enough
$\mathcal{H}$, i.e.~far away from the Feshbach resonance.

In that case, we can expand Eq.~(\ref{eq:fBagnato1}) around the center of mass
$\rho=0$ and $z=z_0$, which gives us in leading order
\begin{equation}
\frac{1}{\mathcal{H} + \frac{M
\omega^2_{\rho}}{2 \mu_{\rm B}}\left(\rho^2+ \lambda^2 z^2 \right)  } \approx
\frac{1}{\mathcal{H} + \frac{M
\omega^2_{\rho}}{2 \mu_{\rm B}}\lambda^2 z_0^2  } \, .
\label{eq:fBagnato2}
\end{equation}
Within this approximation, the integral (\ref{eq:fintegral}) can be evaluated
exactly and yields
\begin{equation}
f(u_{\rho}, u_z, z_0)\approx\frac{\sqrt{2\pi}}{8\mathcal{H}}\frac{ u_{\rho}^2
u_z} {\left(1 + \frac{M \omega_{\rho}^2 \lambda^2 z_0^2}{2 \mu_{\rm B}
\mathcal{H}} \right) } \, .
\label{eq:fBagnato3}
\end{equation}
By substituting Eq.~(\ref{eq:fBagnato3}) into
Eqs.~(\ref{eq:urf})--(\ref{eq:z0f}) and after
introducing dimensionless parameters according to Eq.~(\ref{eq:dimensionlessparameter}) we obtain three second-order ordinary differential equations
for $u_{\rho}$, $u_z$, and $z_0$ \cite{Bagnato12}. A linearization yields the frequencies of collective modes of Ref.~\cite{Bagnato12} in TF approximation to be
\begin{equation}
 \omega^2_{B,Q}= 2 + \frac{3}{2} \lambda^2 \pm \frac{1}{2}\sqrt{16 - 16\lambda^2
+ 9 \lambda^4 }\label{eq:TFFREQ}\, ,
\end{equation}
thus they do not depend on the bias
magnetic field $B_0$. Correspondingly the dipole mode frequency of Ref.~\cite{Bagnato12} in the TF approximation has the form
\begin{align}
\omega^2_D&=\lambda^2 \left(1+\frac{\varepsilon_0
\mathcal{P}_{\rm{BG}}\lambda }{2 \varepsilon_1 u_{\rho 0}^3} \right)\, ,
\label{eq:dipoleBagnatoTF}
\end{align}
where the dipole mode frequency diverges on top of the Feshbach resonance, i.e.~for $\varepsilon_1=0$.
\section{Results}\label{subsec:results}
We discuss in this section the respective results when the bias field $B_0$ is larger than or equal to the
resonant magnetic field $B_{\rm{res}}$, i.e.,  $\mathcal{H}=B_0-B_{\rm{res}}\geq0$. To this end we follow
Refs.~\cite{RB85,RB85Grimm} and consider a concrete experiment with $N=4\times
10^4$ atoms of a ${}^{85}$Rb BEC in a harmonic trap with
$\omega_{\rho}=2 \pi\times156 \ {\rm Hz}$ along the radial direction and
$\omega_z=2 \pi\times16 \ {\rm Hz}$ along the axial direction. The Feshbach resonance
parameters are given by the background value $a_{\rm{BG}}
=-443 a_0$, where $a_0$ is the Bohr radius, the width $\Delta = 10.7$ G, and the
resonance location at $B_{\rm{res}}=155$ G. The magnetic dipole moment $\mu_{\rm B}$ of a
${}^{85}$Rb
\cite{MDipoleMoment} is equal to one Bohr magneton $m_{\rm B}=e \hbar/(2M_e)$, which represents
the magnetic moment of the Hydrogen atom with the elementary charge $e$ and the electron mass $M_e$.
With this the dimensionless parameters (\ref{eq:dimensionlessparameter}) have the values
\begin{equation}
 \mathcal{P}_{\rm{BG}}=-856.732\,, \quad \varepsilon_0\ \varepsilon_1=\varepsilon=9.6052\times 10^4\,.
 \label{eq:Dprameters}
\end{equation}
%

\subsection{Right-Hand Side of Feshbach Resonance}\label{subsec:rightresults}
%
\begin{figure}[!t]
\unitlength1mm
\begin{picture}(90,30)
\put(33,25){\tiny \small(a)}
\put(76,25){\tiny \small(b)}
\put(57,-6){\tiny \small(c)}
\hspace{-0.3cm}\includegraphics[width=4.5cm]{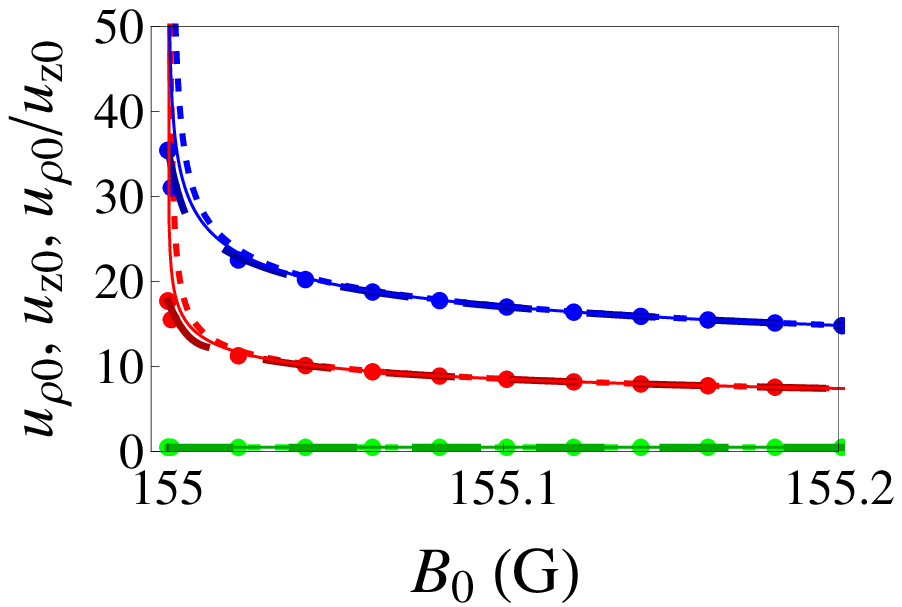}
\hspace{-0.3cm}\includegraphics[width=4.5cm]{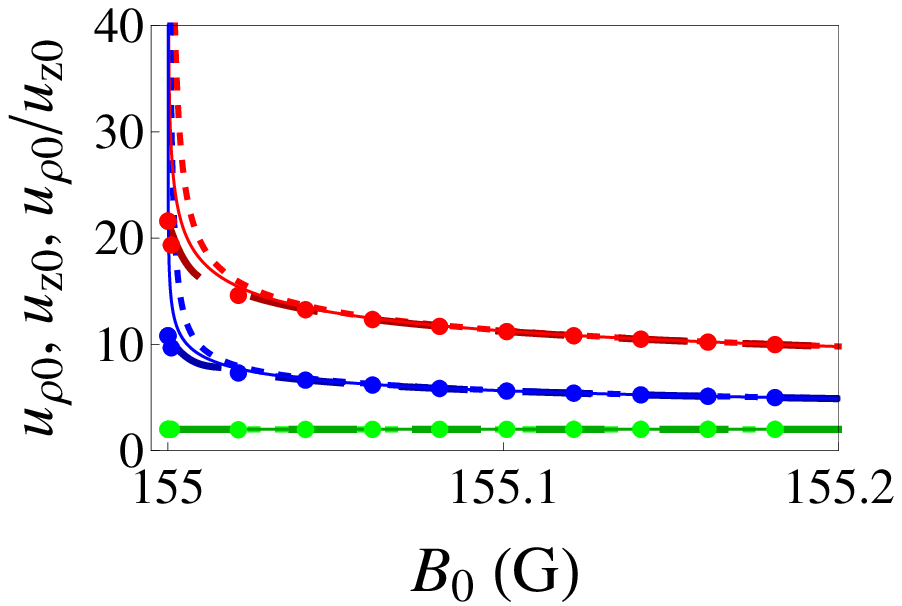}
\end{picture}
\includegraphics[width=4.5cm]{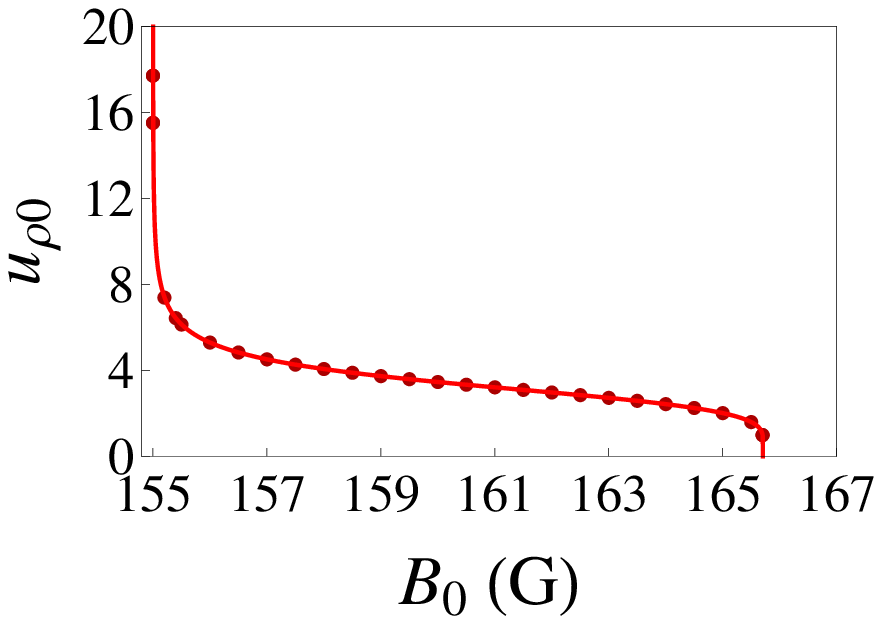}
\vspace{-0.35cm}
\caption{Equilibrium results for condensate widths
$u_{\rho0}$ (red), $u_{z0}$ (blue), and aspect ratio $u_{\rho0}/
               u_{z0}$ (green)  as a function of a magnetic field $B_0$ for
different
trap anisotropy (a), (c) $\lambda=0.5$ and (b) $\lambda=2$ for the experimental
parameters Eq.~(\ref{eq:Dprameters}).
                Solid, dotted, dashed, and square dotted curves correspond to
the heuristic approximation of Ref.~\cite{Bagnato12} and
                the exact results Eqs.~(\ref{eq:ur0B0})--(\ref{eq:uz0B0}) and
the TF approximation Eqs.~(\ref{eq:ur0etaB0}), (\ref{eq:uz0etaB0}), the TF
approximation in the
                limit $\mathcal{H} \rightarrow \infty$ or $\varepsilon_1
\rightarrow \infty$ results Eqs.~(\ref{eq:ur0TFvarepsiloninfty}),
(\ref{eq:uz0TFvarepsiloninfty}), respectively.}
 \label{fig:equlibriumB0}
\end{figure}
We plot in Fig.~\ref{fig:equlibriumB0} the equilibrium widths of the condensate
$u_{\rho0}$, $u_{z0}$, and aspect ratio of $u_{\rho0}/
u_{z0}$  as a function of a magnetic field $B_0$ for the experimental parameters
Eq.~(\ref{eq:Dprameters}) with different trap anisotropy (a), (c) $\lambda=0.5$
and (b)
$\lambda=2$. The widths of the condensate Eqs.~(\ref{eq:ur0B0}) and
(\ref{eq:uz0B0}) are
coupled, so we solve both equations iteratively. We read off that the aspect ratio $u_{\rho0}/u_{z0}$
turns out to coincide perfectly with the trap aspect ratio $\lambda$, therefore, it is justified to use the TF approximation Eq.~(\ref{eq:eta}) to find an analytical understanding for the condensate widths.
From Fig.~\ref{fig:equlibriumB0} we also read off that the heuristic approximation of
Ref.~\cite{Bagnato12} is not valid on top of the
Feshbach resonance and seems to be valid only far away from the Feshbach
resonance. Furthermore, Fig.~\ref{fig:equlibriumB0} confirms that the TF
approximation in Eqs.~(\ref{eq:ur0TF}), (\ref{eq:uz0TF}) agrees quite well with the
equilibrium widths determined from Eq.~(\ref{eq:ur0B0}), (\ref{eq:uz0B0}) as
well as the equilibrium widths calculated
from the limit $\mathcal{H}$ or $\varepsilon_1 \rightarrow \infty$. In addition
Fig.~\ref{fig:equlibriumB0}(c)
shows the radial condensate width $u_{\rho0}$ from Eq.~(\ref{eq:ur0B0}) vanishes
at the critical magnetic field $B_{\rm crit}=B_{\rm res}+\Delta=165.7$ G. As
already anticipated due to a
heuristic argument of Ref.~\cite{Bagnato12}, the system on the right-hand side of
the Feshbach resonance
is not stable beyond this critical magnetic field $B_{\rm crit}$.

\begin{figure}[!t]
\unitlength1mm
\begin{picture}(90,30)
\put(35,22){\tiny \small(a)}
\put(77,22){\tiny \small(b)}
\hspace{-0.3cm}\includegraphics[width=4.45cm]{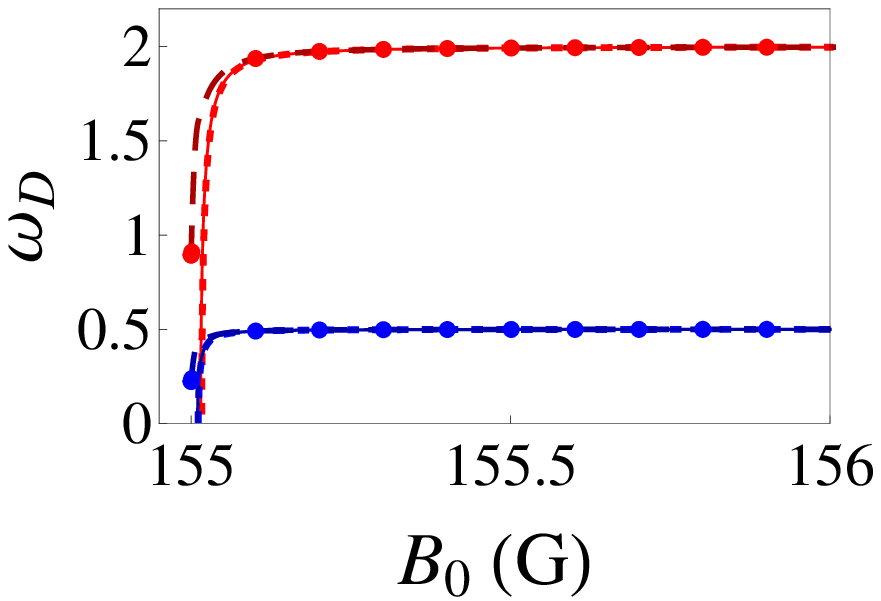}
\hspace{-0.3cm}\includegraphics[width=4.6cm]{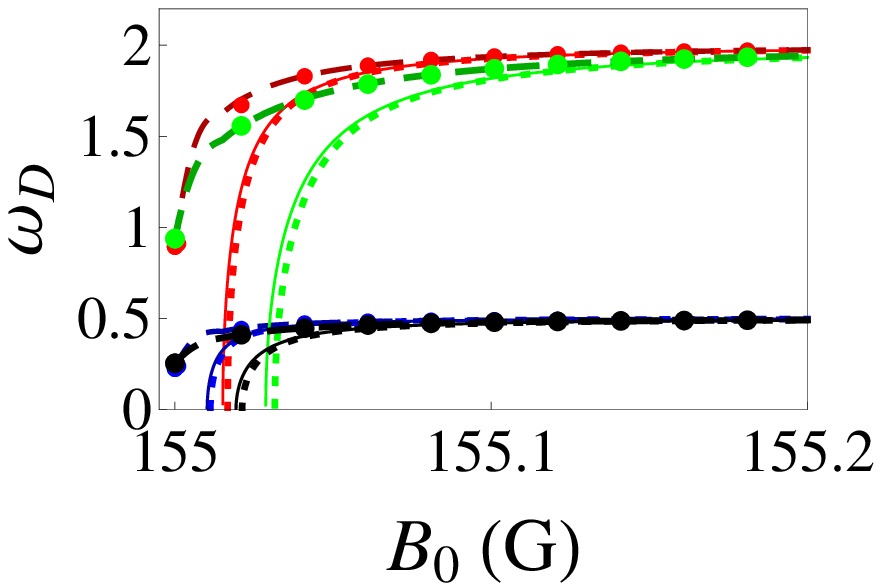}
\end{picture}
\vspace{-0.35cm}
\caption{(a) Dipole mode frequency as a function of a magnetic field
$B_0$ for different trap anisotropy $\lambda=0.5$ (blue) and $\lambda=2$ (red) for the experimental parameters Eq.~(\ref{eq:Dprameters}). Solid, dotted, dashed, and dotted square
curves correspond to the approximation solution of Ref.~\cite{Bagnato12}, the exact result Eq.~(\ref{eq:dipoleB0}) and the TF approximation Eq.~(\ref{eq:dipoleetaB0}) and
the TF approximation in the limit $\mathcal{H} \rightarrow \infty$ Eq.~(\ref{eq:dipoleetaB0varepsiloninfty}), respectively, while
(b) focuses on the region of interest for the dipole mode frequency in addition for the hypothetical value of the Feshbach resonance $\Delta=100.7$ G, with $\lambda=0.5$ (black) and $\lambda=2$ 
(green).}
 \label{fig:dipoleB0}
\end{figure}

\begin{figure}[!t]
\unitlength1mm
\begin{picture}(90,30)
\put(37,26){\tiny \small(a)}
\put(81,26){\tiny \small(b)}
\includegraphics[width=4.5cm]{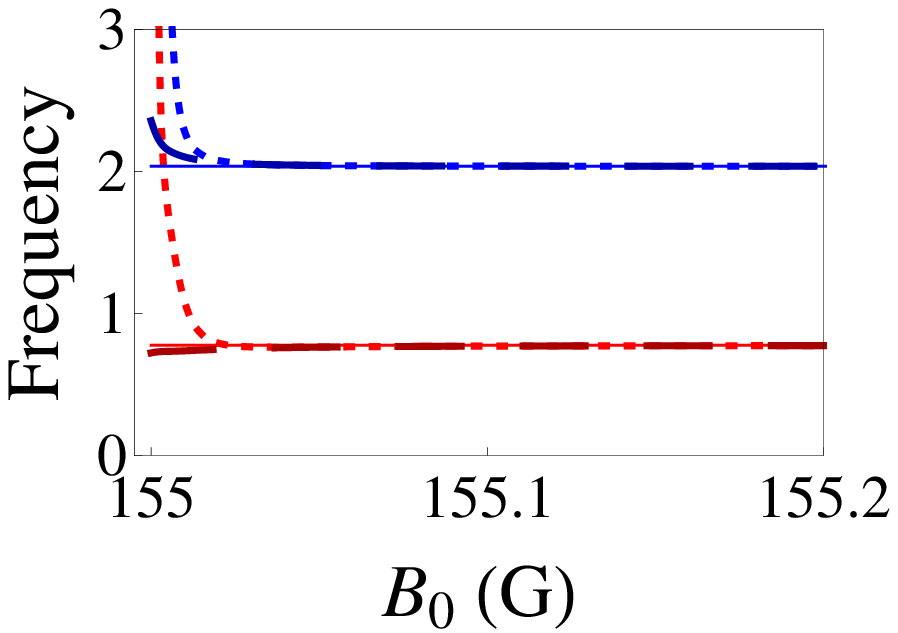}
\hspace{-0.2cm}\includegraphics[width=4.5cm]{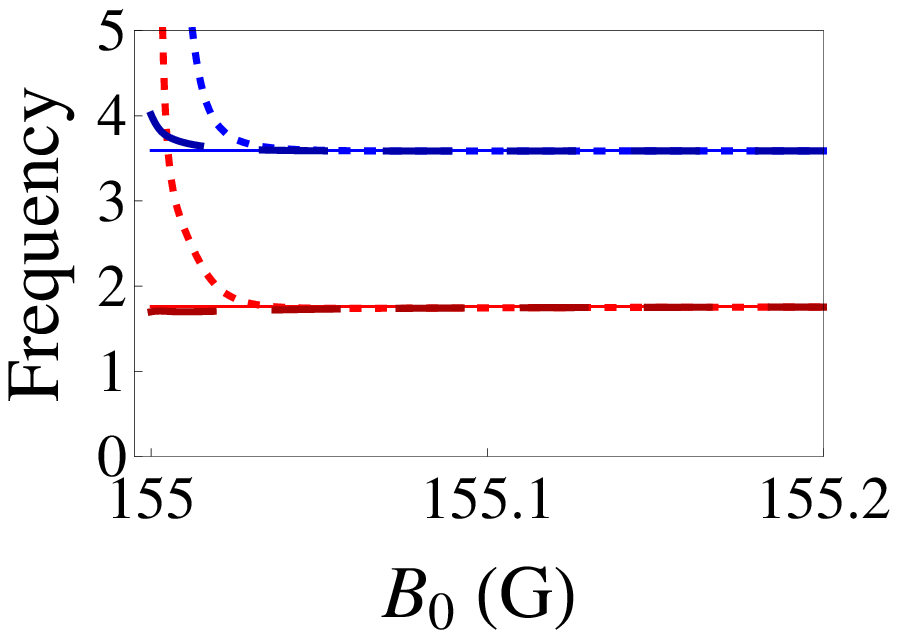}
\end{picture}
\vspace{-0.35cm}
\caption{Frequencies of collective modes results the quadrupole (red) and
breathing (blue) as a function of a magnetic field $B_0$ for different trap
anisotropy (a) $\lambda=0.5$ and (b) $\lambda=2$ for the experimental parameters Eq.~(\ref{eq:Dprameters}). Solid, dashed, and square dotted curves correspond to the approximation solution
of Ref.~\cite{Bagnato12} and the TF approximation
Eq.~(\ref{eq:frequencytop}) with the abbreviations $m_1$, $m_2$, and $m_3$ from Eqs.~(\ref{eq:m1etaB0})--(\ref{eq:m3etaB0}) and the TF approximation in the limit $\mathcal{H}\rightarrow \infty$ or
$\varepsilon_1 \rightarrow \infty$ Eq.~(\ref{eq:frequencytop})
with the abbreviations $m_1$, $m_2$, and $m_3$ being defined in Eqs.~(\ref{eq:m1etaB0varepsiloninfty})--(\ref{eq:m3etaB0varepsiloninfty}), respectively.}
 \label{fig:frequenciesB0}
\end{figure}

Figures \ref{fig:dipoleB0} and \ref{fig:frequenciesB0} show the respective frequencies of
collective modes, for the experimental
parameters Eq.~(\ref{eq:Dprameters}) with different trap anisotropy $\lambda$.
From these figures we see how the frequencies of collective modes change when
one approaches
the Feshbach resonance. As already expected in Eq.~(\ref{eq:omegaeffective}), the dipole mode
frequency on the right-hand side of the Feshbach resonance turns out to be
smaller than the dipole mode frequency far away from the Feshbach resonance. In particular we
observe that the approximative solution of Ref.~\cite{Bagnato12} is not valid on top of
the Feshbach
resonance. Our results and the approximative solution
of Ref.~\cite{Bagnato12} for the dipole mode frequency
in Fig.~\ref{fig:dipoleB0} disagree only $0.05$ G above the Feshbach resonance
for the experimental parameters Eq.~(\ref{eq:Dprameters}). However, this is still
an experimentally accessible range as the magnetic field can be controlled up to
an
accuracy of $1$ mG \cite{Pasquiou}. Furthermore, Fig.~\ref{fig:dipoleB0}(b) shows
how the dipole mode frequency changes with the bias magnetic field $B_0$ for a hypothetical
Feshbach
resonance width $\Delta=100.7$ G. Thus, the difference between
our predication and the approximative solution of Ref.~\cite{Bagnato12} is more
pronounced for a broader Feshbach resonance and for a pancake-like condensate.

\subsection{On Top of Feshbach Resonance}
We remark that approaching
the Feshbach resonance and performing the TF limit represent commuting procedures within our theory.
In contrast to our findings the heuristic approximation of Ref.~\cite{Bagnato12} fails to predict a finite value for the dipole mode frequency
on top of the Feshbach resonance \cite{aljibbouri}.

\begin{figure}[!t]
\unitlength1mm
\begin{picture}(90,30)
\put(37,26){\tiny \small(a)}
\put(52,26){\tiny \small(b)}
\hspace{-0.2cm}\includegraphics[width=4.5cm]{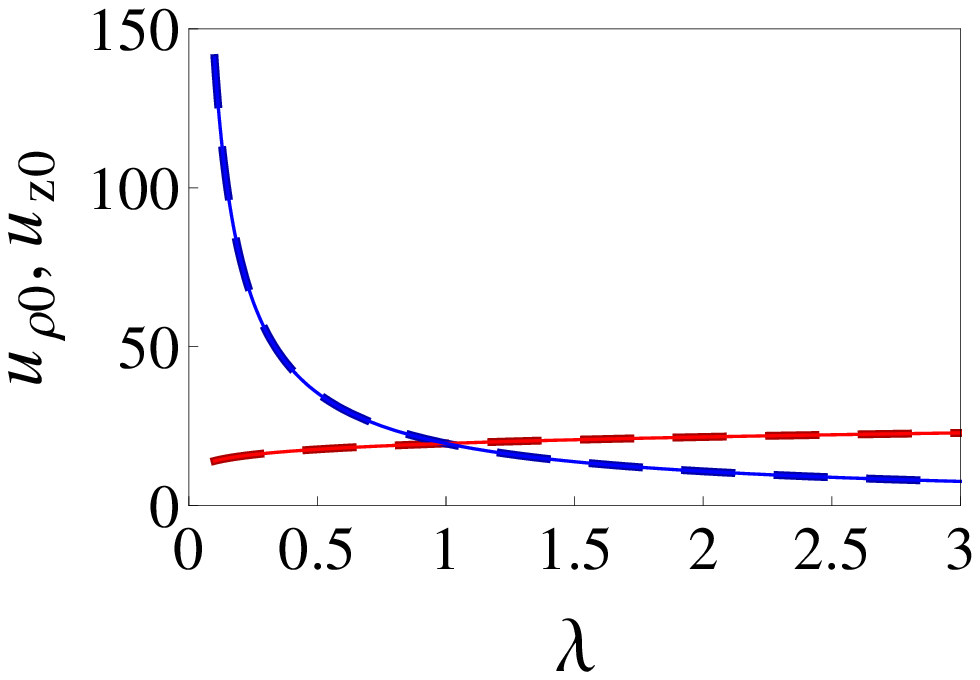}\hspace{0.2cm}
\hspace{-0.2cm}\includegraphics[width=4.3cm]{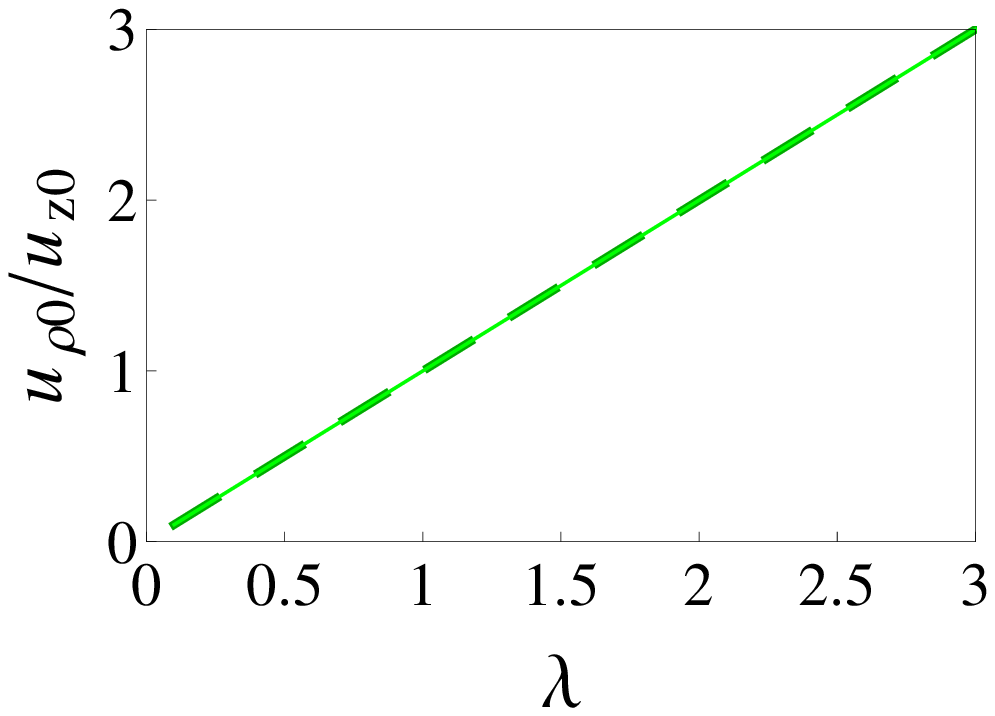}
\end{picture}
\vspace{-0.35cm}
  \caption{Equilibrium results for (a) condensate widths $u_{\rho0}$ (red), $u_{z0}$ (blue) and (b) aspect ratio $u_{\rho0}/u_{z0}$ (green) as a function of trap aspect ratio $\lambda$
               for the experimental parameters Eq.~(\ref{eq:Dprameters}). Solid and dashed curves correspond to the exact results of Ref.~\cite{aljibbouri} and the TF approximation
               Eqs.~(\ref{eq:ur0topTF}), (\ref{eq:uz0topTF}), respectively. }
 \label{fig:Equilibriumtop}
\end{figure}

Figure \ref{fig:Equilibriumtop} shows the equilibrium widths of the condensate $u_{\rho0}$, $u_{z0}$ and the aspect ratio $u_{\rho0}/u_{z0}$ following from the exact results of Ref.~\cite{aljibbouri}
as solid lines versus trap aspect ratio $\lambda$ and the experimental parameters Eq.~(\ref{eq:Dprameters}). From Fig.~\ref{fig:Equilibriumtop}(b) we read off that the aspect ratio $u_{\rho0}/u_{z0}$
turns out to coincide perfectly with the trap aspect ratio $\lambda$.

\begin{figure}[t]
\unitlength1mm
\begin{picture}(90,35)
\put(6,28){\tiny \small (a)}
\put(52.5,28){\tiny \small(b)}
\hspace{-0.2cm}\includegraphics[width=4.5cm]{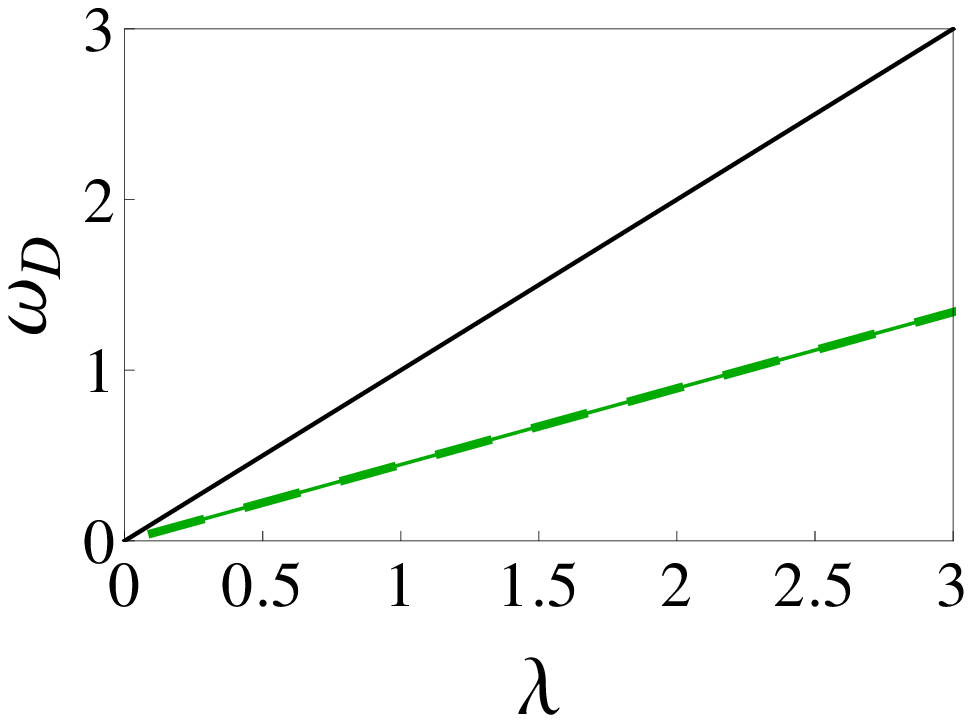}
\includegraphics[width=4.5cm]{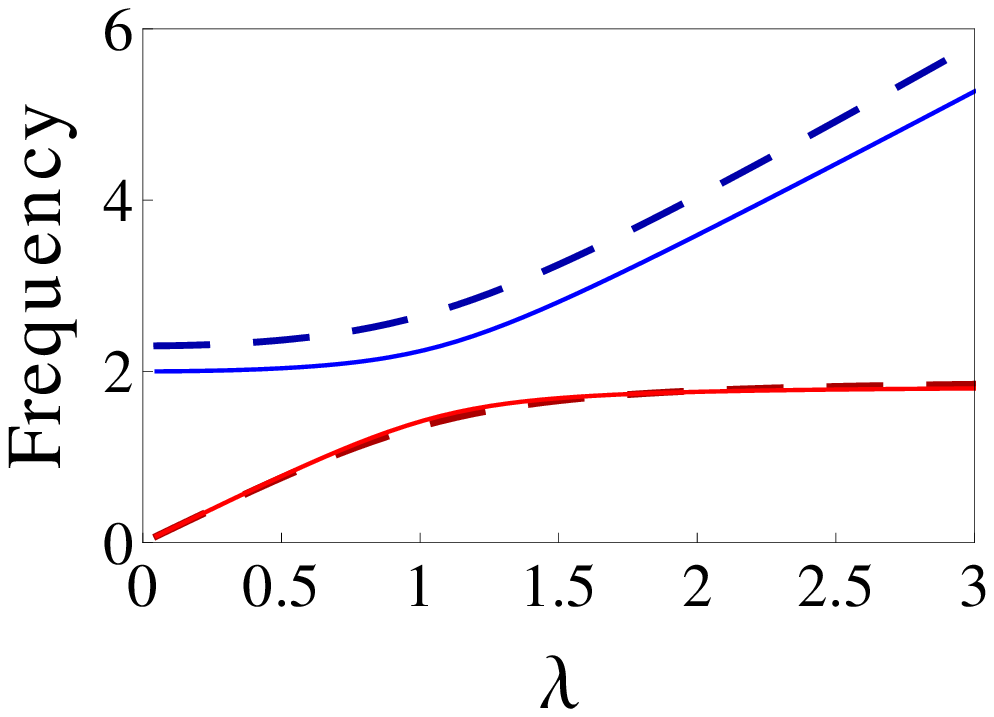}
\end{picture}
\vspace{-0.5cm}
   \caption{Frequencies of collective modes results for (a) the dipole mode frequency (green) and (b) the breathing (blue) and quadrupole (red) mode frequencies as a function
               of trap aspect ratio $\lambda$ for the experimental parameters Eq.~(\ref{eq:Dprameters}). (a) Solid black curve corresponds to the dipole mode frequency far
               away the Feshbach resonance which means that $\omega_D=\lambda$. Solid and dashed curves correspond to the exact result of Ref.~\cite{aljibbouri}
               and in the TF approximation Eq.~(\ref{eq:dipoletopeta}), respectively. (b) Solid and dashed curves correspond to the Eqs.~(\ref{eq:TFFREQ})
               and in the TF approximation Eqs.~(\ref{eq:frequencytop}), where the abbreviations are defined in Eqs.~(\ref{eq:m1eta})--(\ref{eq:m3eta}), respectively.}
 \label{fig:frequencydipoletop}
\end{figure}

In Fig.~\ref{fig:frequencydipoletop}(a) we plot the dipole mode frequency as a function trap
anisotropy $\lambda$. The solid
black curve corresponds to the dipole mode frequency far away from the
Feshbach resonance $\omega_D=\lambda$.
Furthermore, the solid green curve corresponds to the exact result of dipole mode
frequency on top of the Feshbah
resonance \cite{aljibbouri} and the dashed curve corresponds to
the dipole mode in the TF approximation Eq.~(\ref{eq:dipoletopeta})
 for the experimental parameters Eq.~(\ref{eq:Dprameters}).
This result could be seen as being inconsistent with the Kohn theorem \cite{kohn}, which
says that the dipole frequency is equal to the trap frequency and
does not depend on the two-body interaction strength. However, the result of the Kohn
theorem is a consequence of the translational
invariance of the two-body interaction, which is no longer true in our case due to Eq.~(\ref{eq:controlled}). As a consequence the dipole
mode frequency in the exact result of Ref.~\cite{aljibbouri} and its TF approximation
Eq.~(\ref{eq:dipoletopeta}) depend on the two-body interaction strength $\mathcal{P}_{\rm{BG}}$ and 
the anisotropy of the confining potential $\lambda$ both explicitly and implicitly via the equilibrium values of the condensate widths from
 Ref.~\cite{aljibbouri}. 

In Fig.~\ref{fig:frequencydipoletop}(b) we also show the breathing (blue curves) and quadrupole (red curves) mode frequencies as a
function of trap anisotropy $\lambda$. The solid curves correspond to the
frequencies of collective modes far away from the Feshbach resonance, i.e.~for $\varepsilon=0$,
while the
dashed curves correspond to the frequencies of collective mode on top of
the Feshbach resonance and in the TF approximation
Eqs.~(\ref{eq:frequencytop}), with
abbreviations $m_1$, $m_2$, and $m_3$
 are defined in Eqs.~(\ref{eq:m1eta})--(\ref{eq:m3eta}) for the experimental parameters Eq.~(\ref{eq:Dprameters}). We observe that approaching the Feshbach resonance
leads to a significant change of the breathing mode frequency, whereas the quadrupole mode frequency remains basically unaffected.


%
\subsection{Far Away From Feshbach Resonance}
As we have $\varepsilon_0\rightarrow 1/\mathcal{H}$ and $\varepsilon_1 \rightarrow \mathcal{H}$ according to (\ref{eq:dimensionlessparameter}), 
the results
Eqs.~(\ref{eq:ur0TFvarepsiloninfty})--(\ref{eq:dipoleetaB0varepsiloninfty})
represent the $1/\mathcal{H}$ and $1/\mathcal{H}^2$ corrections for the
respective quantities. At first we observe by comparing
Eqs.~(\ref{eq:ur0TFvarepsiloninfty}) and (\ref{eq:uz0TFvarepsiloninfty}) that the heuristic approximation of
Ref.~\cite{Bagnato12} reproduces correctly the $1/\mathcal{H}$ correction for the
condensate widths but fails to determine
the subsequent $1/\mathcal{H}^2$ correction. This is not surprising as the
heuristic approximation (\ref{eq:fBagnato3}) of Ref.~\cite{Bagnato12} for the integral
(\ref{eq:fintegral}) is only exact up to order $1/\mathcal{H}$. But we read off from our results  in Eq.~(\ref{eq:dipoleetaB0varepsiloninfty})
for the dipole mode frequency, plotted in Fig.~\ref{fig:dipoleB0}, that the
leading order correction to the Kohn theorem near Feshbach resonance is in fact of the
order $1/\mathcal{H}^2$. Therefore, the corresponding predication
of the heuristic approximation of Ref.~\cite{Bagnato12} is even incorrect far
away from the Feshbach resonance.

In addition the similar situation for the breathing and quadrupole frequencies
shows that the leading order of our results (\ref{eq:frequencytop}), with
the abbreviations $m_1$, $m_2$, and $m_3$
from
Eqs.~(\ref{eq:m1etaB0varepsiloninfty})--(\ref{eq:m3etaB0varepsiloninfty}), presented in
Fig.~\ref{fig:frequenciesB0}, is $1/\mathcal{H}^2$ and that the frequencies
depend strongly on the magnetic field $B_0$ and are divergent on top of
the Feshbach resonance,
while the frequencies of the heuristic approximation of Ref.~\cite{Bagnato12} fail to determine the correct  $1/\mathcal{H}^2$ correction and depend only
on the trap anisotropy $\lambda$, i.e., they do not depend on the bias magnetic field
$B_0$.

\section{Conclusions}\label{sec:CONCLUSIONSS}
We have studied in detail how the dipole mode frequency and the collective excitation modes of a
harmonically trapped Bose-Einstein condensate plus a bias potential change on the right-hand side and on top of the Feshbach resonance.
To this end, we have derived equations of motion (\ref{eq:urf})--(\ref{eq:z0f}) for the variational parameters which describe the  radial and
axial condensate widths as well as the center of mass position and
have shown 
how to extract the frequencies of the low-lying collective modes. At first we have analyzed our own treatment which is based on rewriting the integral in Eq.~(\ref{eq:fintegral1})
with the help of the Schwinger trick \cite{schwinger}. Then we have studied the consequences
of this integral representation for the collective mode frequencies both on the right-hand side and on top of the Feshbach resonance.

On the right-hand side of the Feshbach
resonance we found that the system is not stable beyond the critical magnetic field $B_{\rm crit}$. Furthermore, we have shown how
the frequencies of the collective modes change when one approaches the Feshbach resonance. As expected initially the dipole mode frequency for the exact result and TF
approximation on the right-hand side of the
Feshbach resonance turn out to be smaller than the dipole mode far away from the Feshbach resonance.
Furthermore we discussed the TF approximation for the condensate widths and the frequencies of collective modes in two limits. At first we considered the limit on top of the
Feshbach resonance, i.e.~$\mathcal{H}\rightarrow 0$ or $\varepsilon_1 \rightarrow 0$, and afterwards, we discussed the limit far away from the Feshbach
resonance, i.e.~$\mathcal{H}\rightarrow \infty$ or $\varepsilon_1 \rightarrow \infty$.

Our results and the approximative solution of Ref.~\cite{Bagnato12} disagree for only about $0.05$ G above the Feshbach resonance for the experimental parameters
of Refs.~\cite{RB85,RB85Grimm}, but this is still large enough to be experimentally accessible as the magnetic field can be tuned up to $1$ mG \cite{Pasquiou}. 
Thus, the presented results for the violation of the Kohn theorem could, in principle, be detected in future experiments.  It would be interesting to study how these results change by taking 
into account quantum fluctuations \cite{Arstieu1,Arstieu2}.

\section*{Acknowledgments}
We thank Vanderlei Bagnato, Antun Bala\v{z}, and Ednilson Santos for
inspiring discussions. Furthermore we acknowledge
financial support from the German Academic Exchange
Service (DAAD) as well as from the
German Research Foundation (DFG) via the Collaborative
Research Center SFB/TR49 Condensed Matter Systems with Variable Many-Body
Interactions.


\begin{thebibliography}{100}
\bibitem{StamperKurn}D. M. Stamper-Kurn, H. J. Miesner, S. Inouye, M. R.
Anderws, and
W. Ketterle, Phys. Rev. Lett. \textbf{81}, 500 (1998).

\bibitem{DSJin} D. S. Jin, J. R. Ensher, M. R. Matthews, C. E. Wieman, and E. A.
Cornell, Phys. Rev. Lett. {\bf 77}, 420 (1996).

\bibitem{MOMewes} M.-O. Mewes, M. R. Andrews, N. J. van Druten, D. M. Kurn, D.
S.
Durfee, C. G. Townsend, and W. Ketterle, Phys. Rev. Lett. {\bf 77}, 988 (1996).

\bibitem{YCastinandRDum} Y. Castin and R. Dum, Phys. Rev. Lett. {\bf 77}, 5315
(1996).

\bibitem{FDalfovo}  F. Dalfovo, C. Minniti, and L. P. Pitaevskii, Phys. Rev.
A. {\bf 56}, 4855 (1997).

\bibitem{JJGarciaRipoll} J. J. Garc\'ia-Ripoll, V. M. P\'erez-Garc\'ia, and P.
Torres, Phys.
Rev. Lett. {\bf 83}, 1715 (1999).

\bibitem{JJGarciaRipoll1} J. J. G. Ripoll and V. M. P\'erez-Garc\'ia,
Phys. Rev. A {\bf
59}, 2220 (1999).

\bibitem{AINicolin} A. I. Nicolin, Phys. Rev. E {\bf 84}, 056202 (2011).

\bibitem{GHechenblaikner1}  G. Hechenblaikner, O. M. Marag\`o, E. Hodby, J.
Arlt, S. Hopkins,
and C. J. Foot, Phys. Rev. Lett. {\bf 85}, 692 (2000).

\bibitem{hodby} E. Hodby, O.M. Marag\`{o}, G. Hechenblaikner, and C.J. Foot,
Phys. Rev. Lett. {\bf 86}, 2196 (2001).

\bibitem{YuZhou} 
Y. Zhou, W. Wen, and G. Huang, Phys Rev. B {\bf 77}, 104527 (2008).

\bibitem{Hamid1}I. Vidanovi\'c, H. Al-Jibbouri, A. Bala\v{z}, and A. Pelster,
Phys. Scr. T {\bf 149}, 014003 (2012).

\bibitem{Hamid2} H. Al-Jibbouri, I. Vidanovi\'c, A. Bala\v{z}, and A. Pelster,
J. Phys. B {\bf 46}, 065303 (2013).

\bibitem{VSBagnato0}
E. R. F. Ramos, E. A. L. Henn, J. A. Seman, M. A. Caracanhas, K. M. F. Magalh\~aes, 
K. Helmerson, V. I. Yukalov, and V. S. Bagnato, Phys. Rev. A {\bf 78}, 063412 (2008).

\bibitem{SEPollackandRGHulet} S. E. Pollack, D. Dries, M. Junker, Y. P. Chen, T.
A. Corcovilos,
and R. G. Hulet,
 Phys. Rev. Lett. {\bf 102}, 090402 (2009).

\bibitem{VSBagnato} S. E. Pollack, D. Dries, R. G. Hulet, K. M. F. Magalh\~aes,
E. A.
L.
Henn, E. R. F. Ramos, M. A. Caracanhas, and
V. S. Bagnato, Phys. Rev. A {\bf 81}, 053627 (2010).

\bibitem{IVidanovic} I. Vidanovi\'c, A. Bala\v{z}, H. Al-Jibbouri, and A. Pelster,
Phys. Rev. A {\bf 84}, 013618 (2011).

\bibitem{SSabari} S. Sabari, R. V. J. Raja, K. Porsezian and P.
Muruganandam, J. Phys. B: At. Mol. Opt. Phys. {\bf 43}, 125302 (2010).

\bibitem{AINicolin1} A. I. Nicolin, Rom. Rep. Phys., {\bf 63}, 1329
(2011).

\bibitem{will}W. Cairncross and A. Pelster, arXiv:1209.3148.

\bibitem{kohn}W. Kohn, Phys. Rev. {\bf 123}, 1242 (1961).

\bibitem{DanielRokhsar}A. L. Fetter and D. Rokhsar, Phys. Rev. A {\bf 57}, 1191
(1998).

\bibitem{Zaremba2771999}E. Zaremba, T. Nikuni, and A. Griffin, J. Low Temp.
Phys. {\bf 116}, 277 (1999).

\bibitem{STOOFkohn}M. J. Bijlsma and H. T. C. Stoof, Phys. Rev. A {\bf 60}, 3973
(1999).

\bibitem{Minguzzi}A. Minguzzi and M. P. Tosi, J. Phys.: Condens. Matter {\bf 9},
10211 (1997).

\bibitem{Jurgen}J. Reidl, G. Bene, R. Graham, and P. Sz\'epfalusy, Phys. Rev. A
{\bf 63}, 043605 (2001).

\bibitem{sum1}A. Minguzzi, Phys. Rev. A {\bf 64}, 033604 (2001).

\bibitem{sum11}T. Maruyama and G. F. Bertsch, Phys. Rev. A {\bf 77}, 063611
(2008).

\bibitem{sum2}A. Banerjee, J. Phys. B: At. Mol. Opt. Phys. {\bf 42},
 235301 (2009).

\bibitem{sum3}J.-Y. Zhang, S.-C. Ji, Z. Chen, L. Zhang, Z.-D. Du, B. Yan, G.-S.
Pan, B. Zhao, Y.-J. Deng, H. Zhai, S. Chen, and J.-W. Pan,
Phys. Rev. Lett. {\bf 109}, 115301 (2012).

\bibitem{sum4}Y. Li, G. I. Martone, and S. Stringari, Euro. Phys. Lett. {\bf
99}, 56008 (2012).

\bibitem{sum5} F. Ferlaino, R. J. Brecha, P. Hannaford, F. Riboli, G. Roati, G.
Modugno, and M. Inguscio, J. Opt. B: Quantum Semiclass. Opt. {\bf 5}, S3 (2003).

\bibitem{sum6} S. Chiacchiera, T. Macr\`\i and A. Trombettoni, Phys. Rev. A {\bf
81}, 033624 (2010).

\bibitem{YANGLu} Y. Lu, W. Xiao-Rui, L. Ke, T. Xin-Zhou, X. Hong-Wei, and L.
Bao-Long, Chin. Phys. Lett. {\bf 26}, 076701 (2009).

\bibitem{SStringari} S. Stringari, Phys. Rev. Lett. {\bf 77}, 2360 (1996).

\bibitem{CJPethick} C.~J. Pethick and H. Smith, \emph{Bose-Einstein Condensation
in Dilute Gases}, 2nd edition (Cambridge University Press, Cambridge, 2008).

\bibitem{HOtt}H. Ott, J. Fort\'agh, S. Kraft, A. G\"unther, D. Komma, and
C. Zimmermann, Phys. Rev. Lett. {\bf 91}, 040402 (2003).

\bibitem{Bagnato12} E. R. F. Ramos, F. E. A. dos Santos, M. A. Caracanhas, and
V. S. Bagnato, Phys. Rev. A {\bf 85}, 033608 (2012).

\bibitem{Ioffe}T. Esslinger, I. Bloch, and T. W. Hansch, Phys. Rev. A {\bf 58}, R2664 (1998).

\bibitem{RB85}P. A. Altin, N. P. Robins, D. D\"oring, J. E. Debs, R. Poldy, C.
Figl, and J. D. Close, Rev. Sci. Instrum. {\bf 81}, 063103 (2010).

\bibitem{RB85Grimm}C. Chin, R. Grimm, P. Julienne, and E. Tiesinga, Rev. Mod.
Phys. {\bf 82}, 1225 (2010).

\bibitem{Grosseq}E. P. Gross, Nuovo Cimento, {\bf 20}, 454 (1961).

\bibitem{Pitaevskiieq} 
L. P. Pitaevskii, Sov Phys. JETP {\bf 13}, 451 (1961).

\bibitem{Perezandzoller} 
V. M. P\'erez-Garc\'\i a, H. Michinel, J . I. Cirac, M. Lewenstein,
and P. Zoller, Phys. Rev. Lett. {\bf 77}, 5320
(1996).

\bibitem{VMPerez}V. M. P\'erez-Garc\'\i a, H. Michinel, J. I. Cirac, M.
Lewenstein, P.
Zoller,
Phys. Rev. A {\bf 56}, 1424 (1997).

\bibitem{schwinger} H. Kleinert and V. Schulte-Frohlinde, {\it Critical
Properties of $\Phi^4$-Theories} (World Scientific, Singapore, 2001).

\bibitem{Moerdijk}A. J. Moerdijk, B. J. Verhaar, and A. Axelsson, Phys. Rev. A
{\bf 51},
4852 (1995).

\bibitem{MDipoleMoment}S. Yi and L. You, Phys. Rev. A {\bf 67}, 045601
(2003).

\bibitem{Mathematica} {\it Mathematica} symbolic calculation software package,
\url{http://www.wolfram.com/mathematica}.

\bibitem{Pasquiou}B. Pasquiou, E. Mar\'echal, G. Bismut, P. Pedri, L. Vernac, O. Gorceix, and B. Laburthe-Tolra, Phys. Rev. Lett. {\bf 106}, 255303 (2011).

\bibitem{aljibbouri}H. Al-Jibbouri, Collective Excitations in Bose-Einstein Condensates, PhD thesis, Free University of Berlin, In preparation.

\bibitem{Arstieu1}A. R. P. Lima and A. Pelster, Phys. Rev. A {\bf 84}, 041604 (2011).

\bibitem{Arstieu2}A. R. P. Lima and A. Pelster, Phys. Rev. A {\bf 86}, 063609 (2012).

\end{thebibliography}
\end{document}